%% file: iet_software2009.tex
\documentclass{article}

\usepackage{graphicx,amsthm,amsmath,amssymb}

\begin{document}
\title{The performance of locality-aware topologies for peer-to-peer live streaming}

\author{
    R.~G.~Clegg (richard@richardclegg.org) \and
    R.~Landa (rlanda@ee.ucl.ac.uk) \and    
    D.~Griffin (dgriffin@ee.ucl.ac.uk) \and
    E.~Mykoniati (e.mykoniati@ee.ucl.ac.uk) \and 
    M.~Rio(m.rio@ee.ucl.ac.uk) \and
}
\date{}

\newtheorem{theorem}{Theorem}
\theoremstyle{definition}
\newtheorem{definition}{Definition}
\newtheorem*{remark}{Remark}

\newcommand{\mr}{\mathbb{R}} 
\newcommand{\mn}{\mathbb{N}} 
\newcommand{\mz}{\mathbb{Z}} 
\newcommand{\var}[1]{\text{var}\left(#1\right)}
\newcommand{\Esym}{\mathrm{E}}
\newcommand{\E}[1]{\Esym\left[#1\right]}
\newcommand{\Probsym}{\mathbb{P}}
\newcommand{\Prob}[1]{\Probsym\left[#1\right]}
\newcommand{\bD}{\mathbf{D}}
\newcommand{\bP}{\mathbf{P}}
\newcommand{\bV}{\mathbf{V}}
\newcommand{\bS}{\mathbf{S}}
\newcommand{\cN}{\mathcal{N}}
\newcommand{\cP}{\mathcal{P}}
\newcommand{\cB}{\mathcal{B}}
\newcommand{\cV}{\mathcal{V}}

\maketitle
\begin{abstract}
This paper is concerned with the effect of overlay network topology on the performance 
of live streaming peer-to-peer systems. The paper focuses on the evaluation of topologies which
are aware of the delays experienced between different peers on the network.
Metrics are defined which assess the topologies
in terms of delay, bandwidth usage and resilience to peer drop-out.  Several topology creation algorithms
are tested and the metrics are measured in a simple simulation
testbed.  This gives an assessment of the type of gains which might be
expected from locality awareness in peer-to-peer networks.
\end{abstract}

\section{Introduction}
\label{sec:intro}
\input{introduction}

\section{Simulation method}
\label{sec:method}
\input{method}

\section{Topology construction and metrics}
\label{sec:topol}
\input{topologies}

\section{Results}
\label{sec:results}
\input{results}

\section{Conclusions and further work}
\label{sec:conc}
\input{conclusions}

\bibliographystyle{ieeetr}
\bibliography{iet_software2009}

\pagebreak

\begin{figure}[ht!]
\begin{center}
\includegraphics[width=6cm]{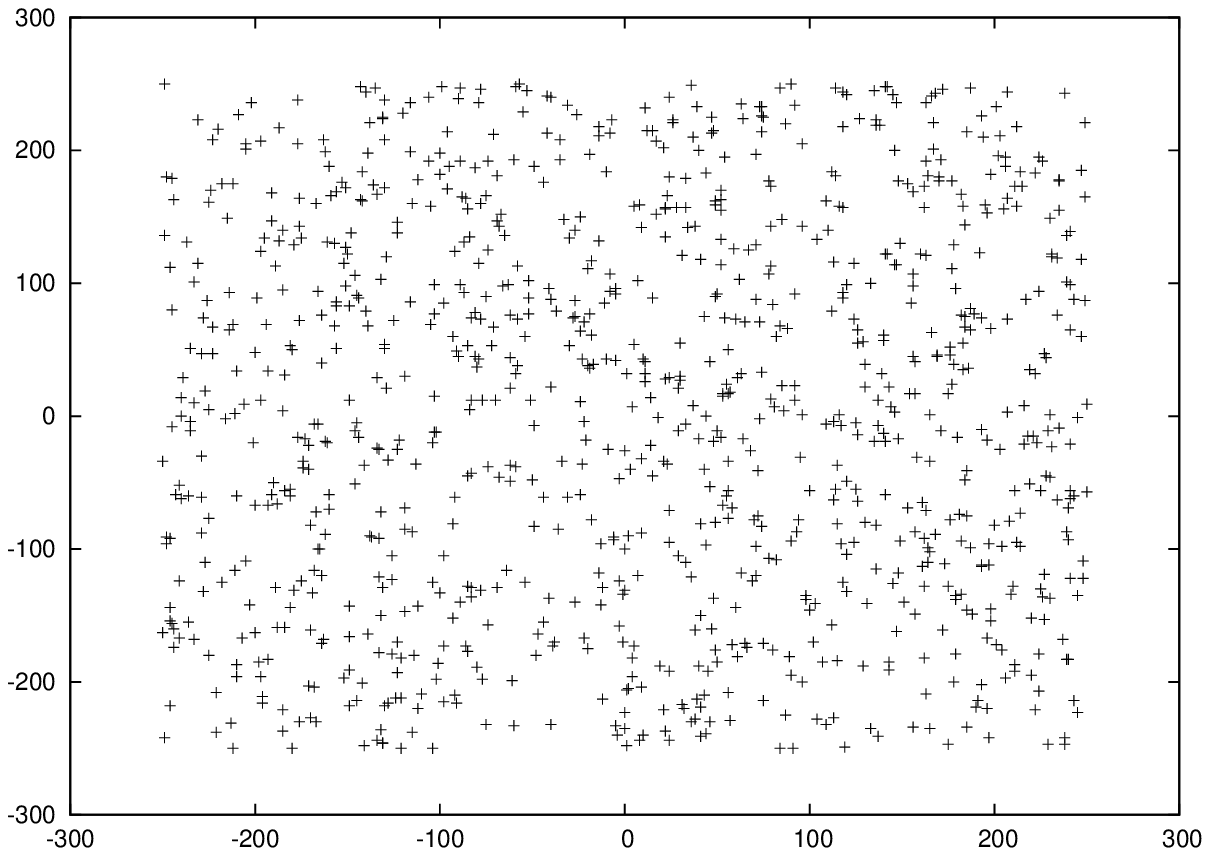}
\includegraphics[width=6cm]{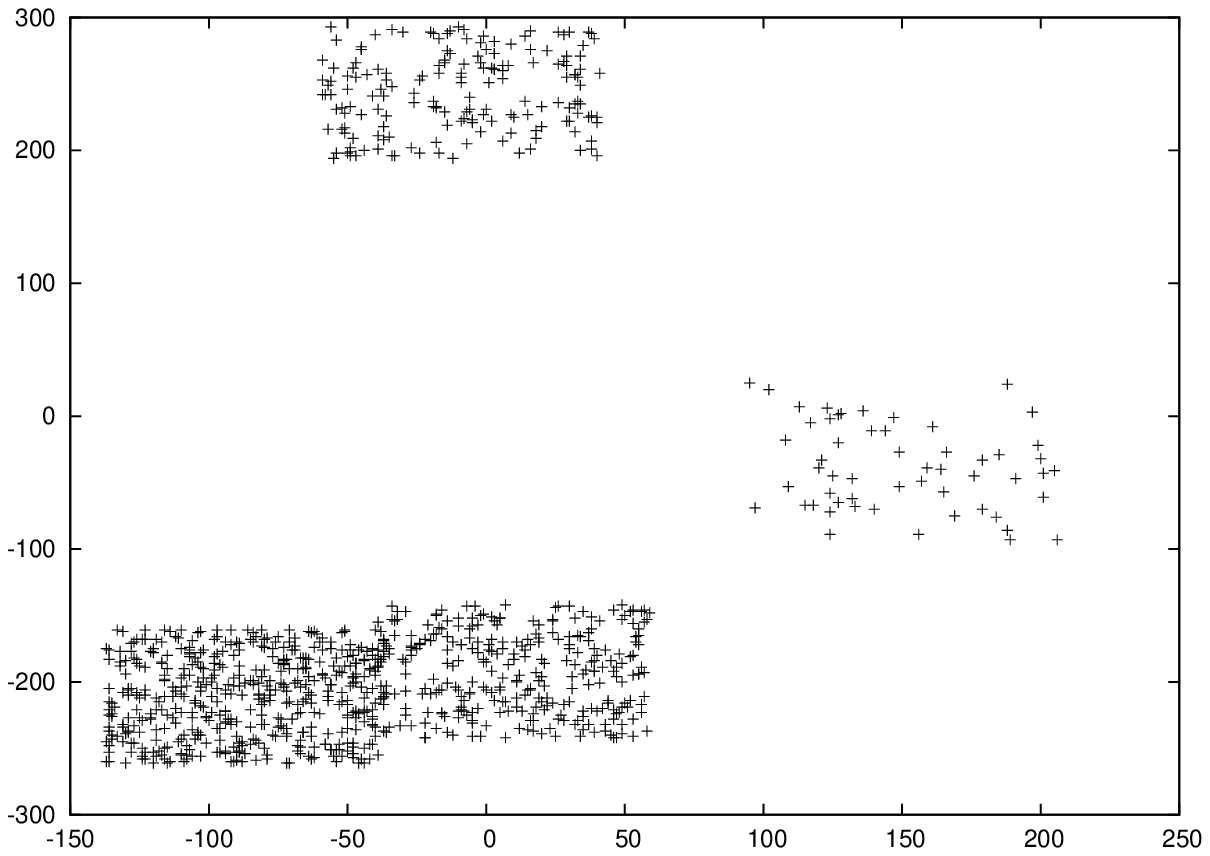}
\caption{Flat (left) and loosely clustered (right) node distributions
of 1000 nodes.}
\label{fig:distributions}
\end{center}
\end{figure}

\begin{figure}
\begin{center}
\includegraphics[width=6cm]{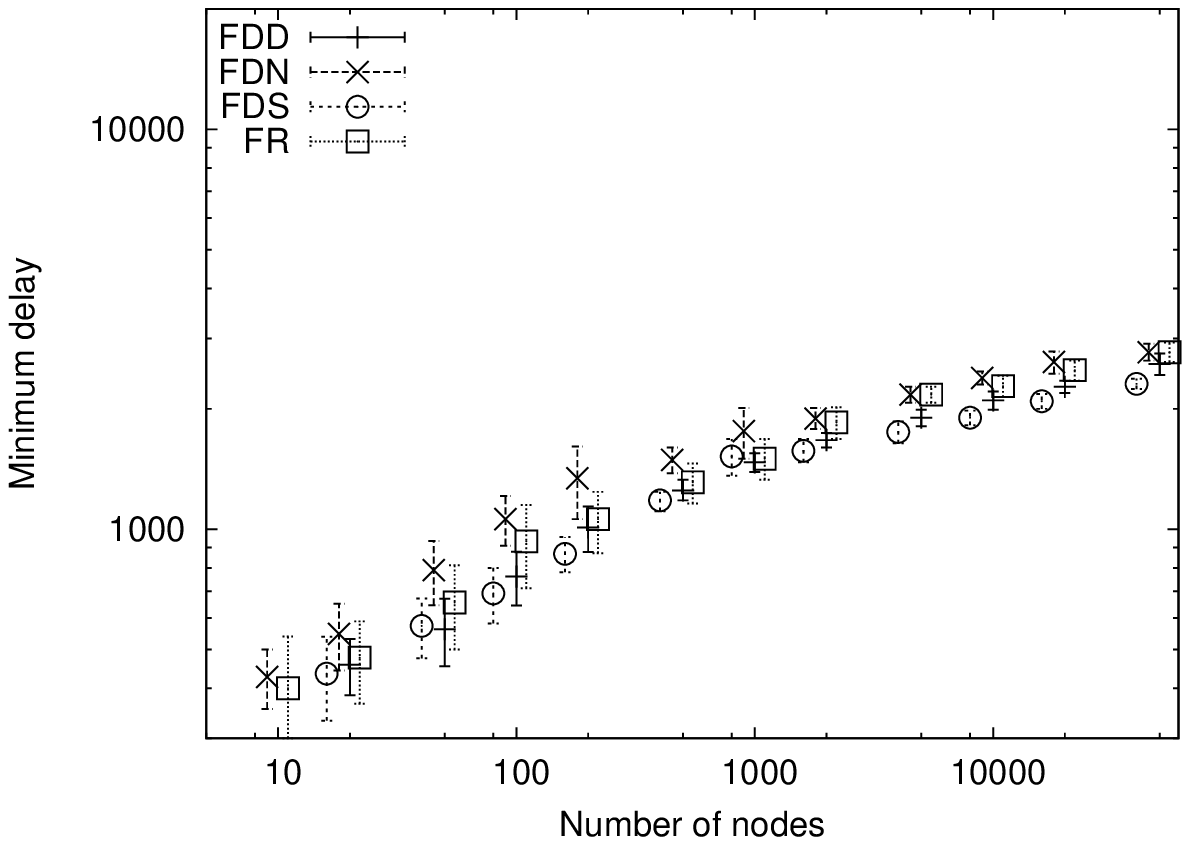}
\includegraphics[width=6cm]{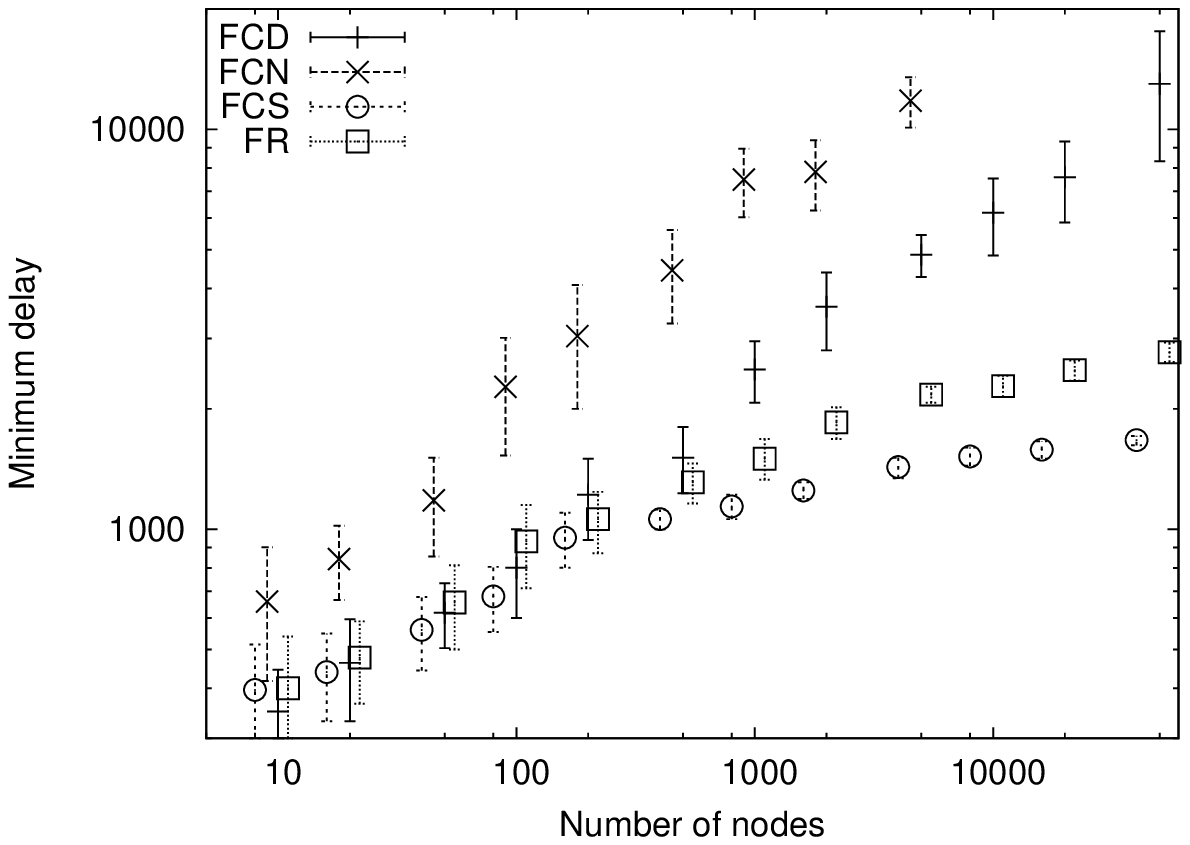}\newline
\includegraphics[width=6cm]{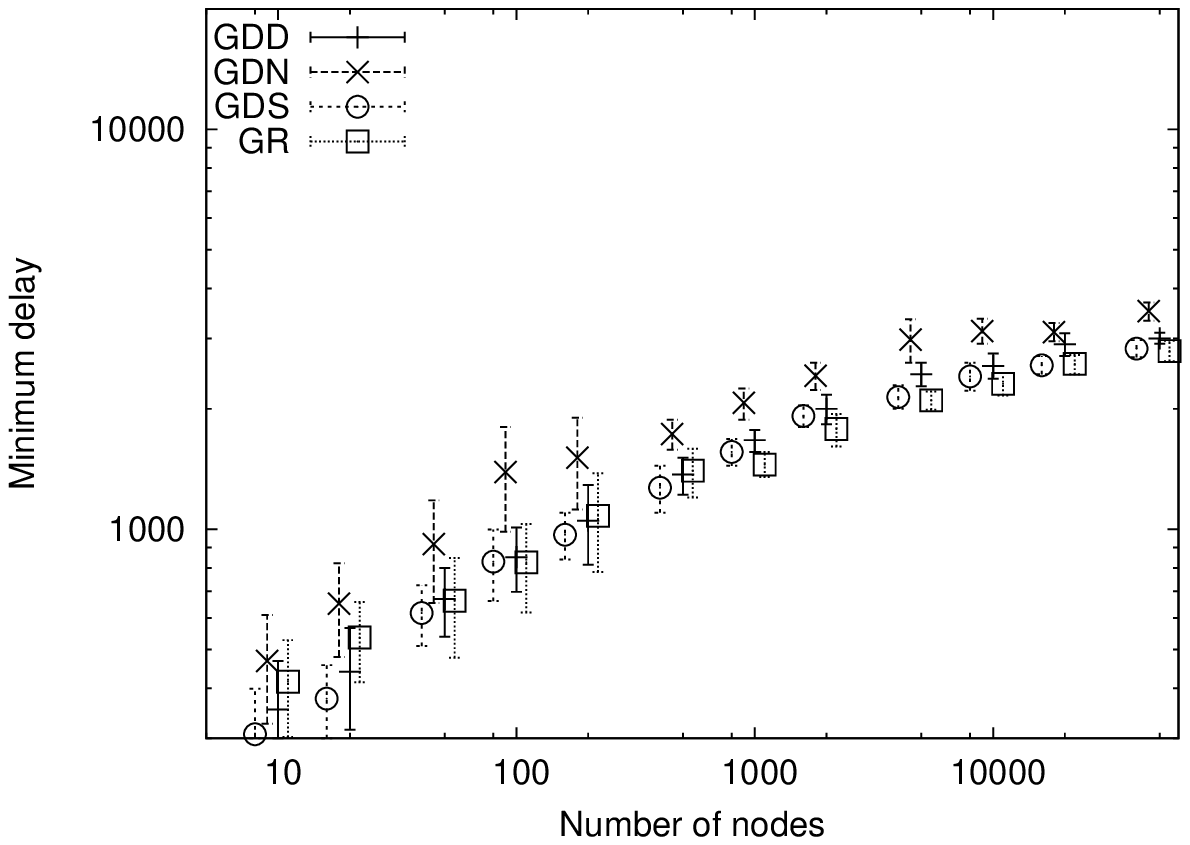}
\includegraphics[width=6cm]{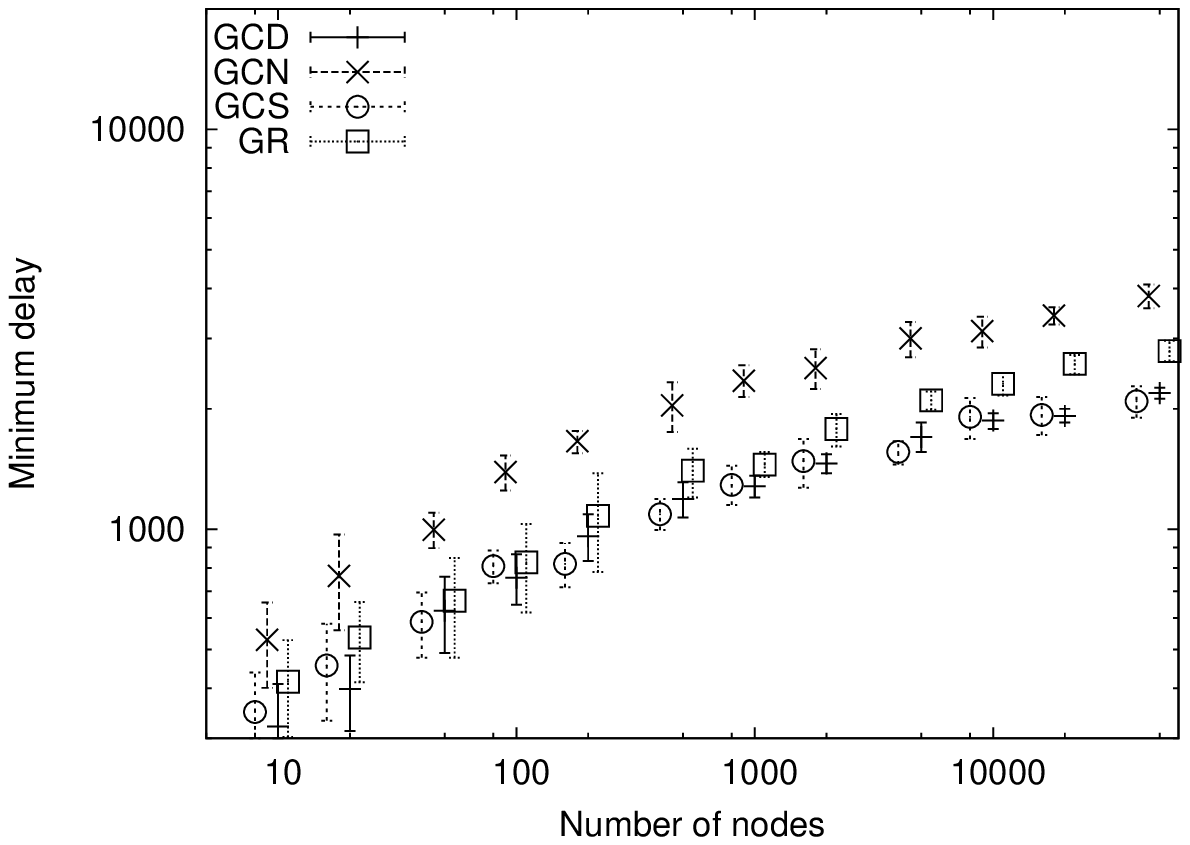}\newline
\caption{Minimum delay for the various topologies and all node distributions.}
\label{fig:del}
\end{center}
\end{figure}

\begin{figure}
\begin{center}
\includegraphics[width=6cm]{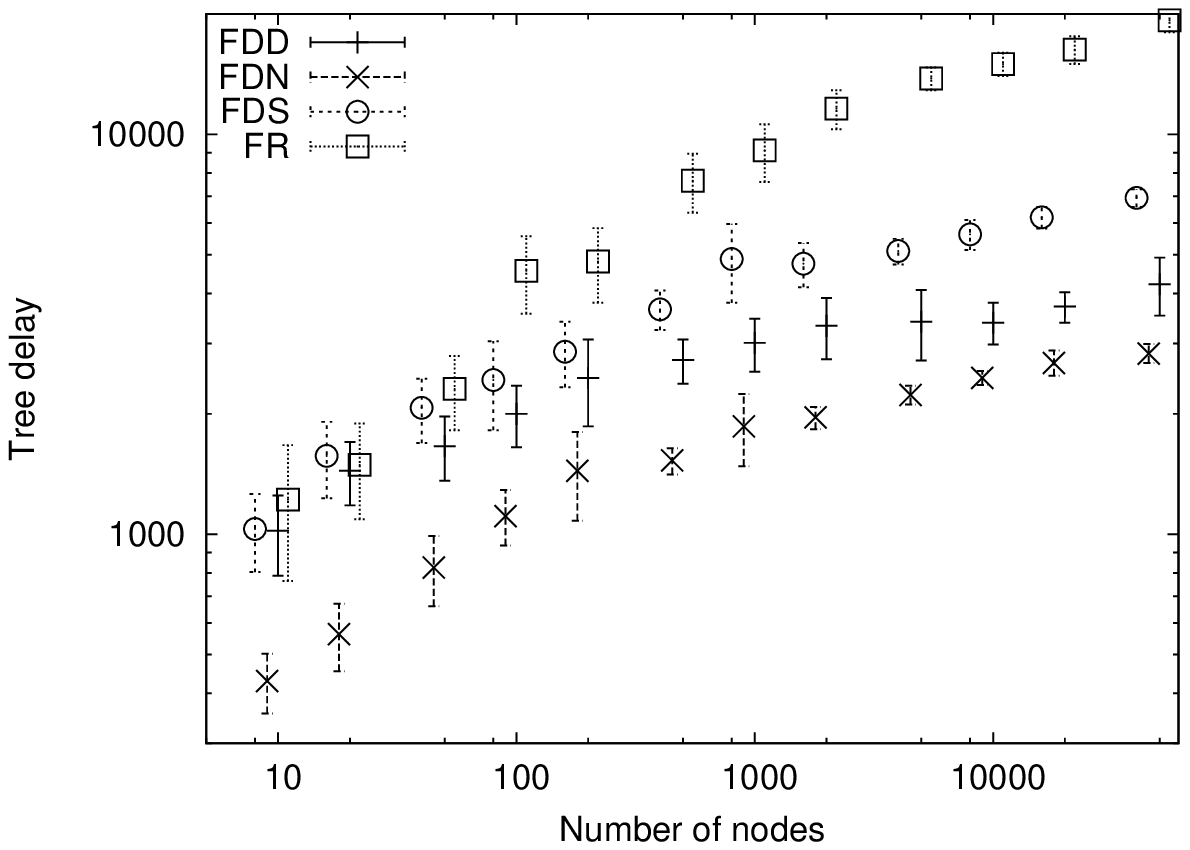}
\includegraphics[width=6cm]{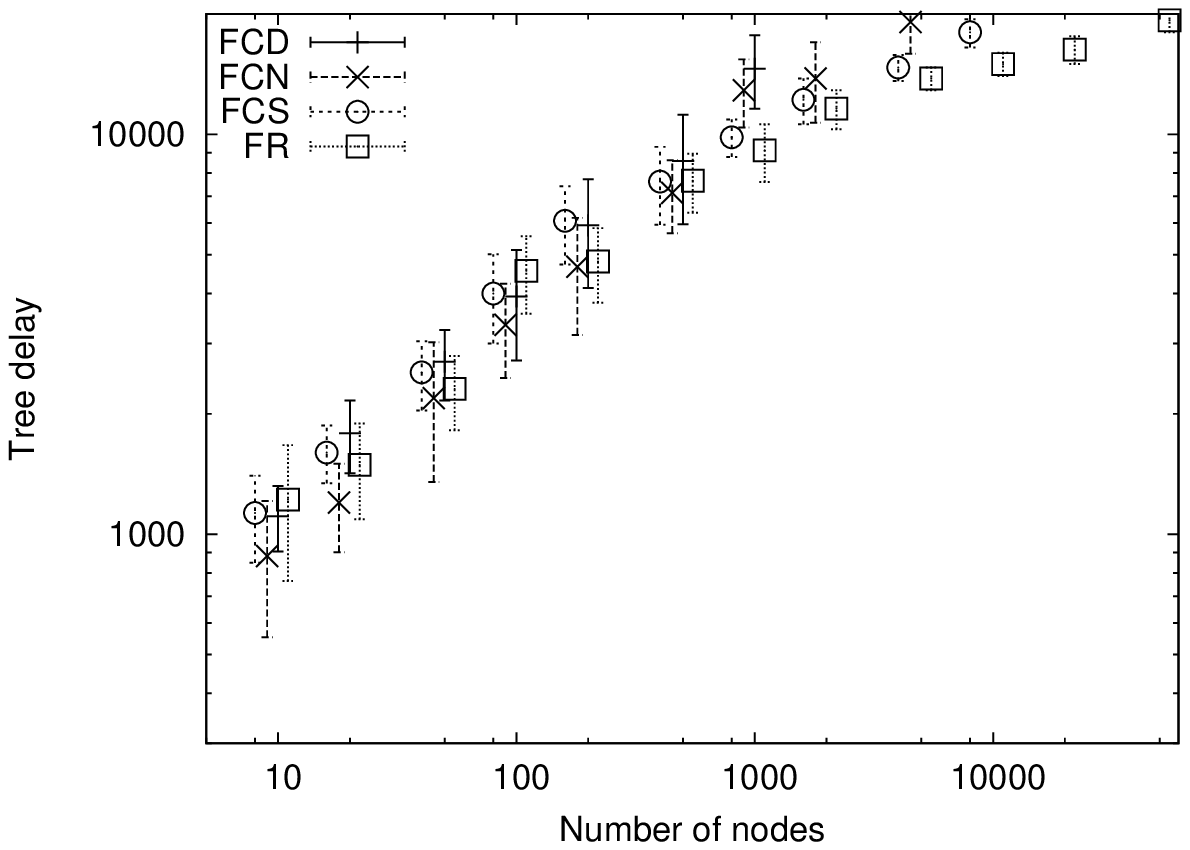}\newline
\includegraphics[width=6cm]{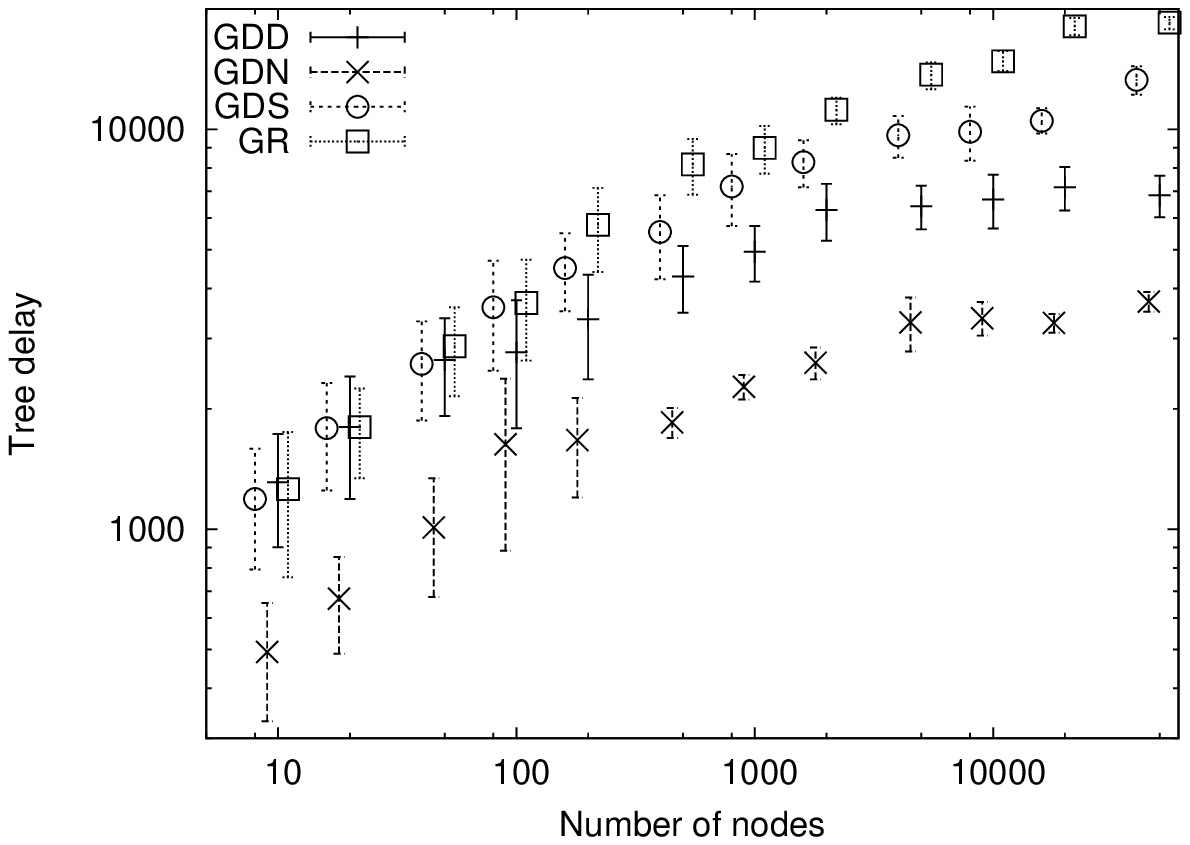}
\includegraphics[width=6cm]{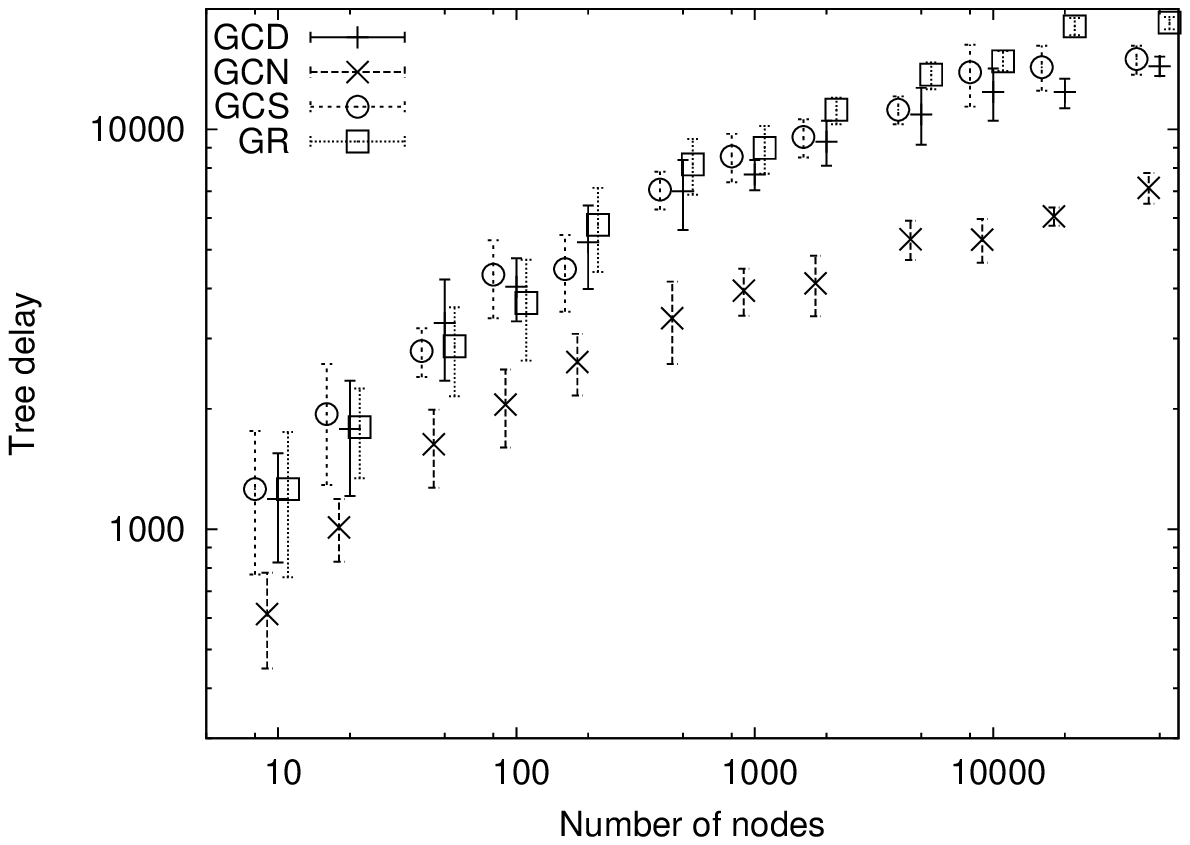}\newline
\caption{Tree delay for the various topologies and all node distributions.}
\label{fig:tree}
\end{center}
\end{figure}

\begin{figure}
\begin{center}
\includegraphics[width=6cm]{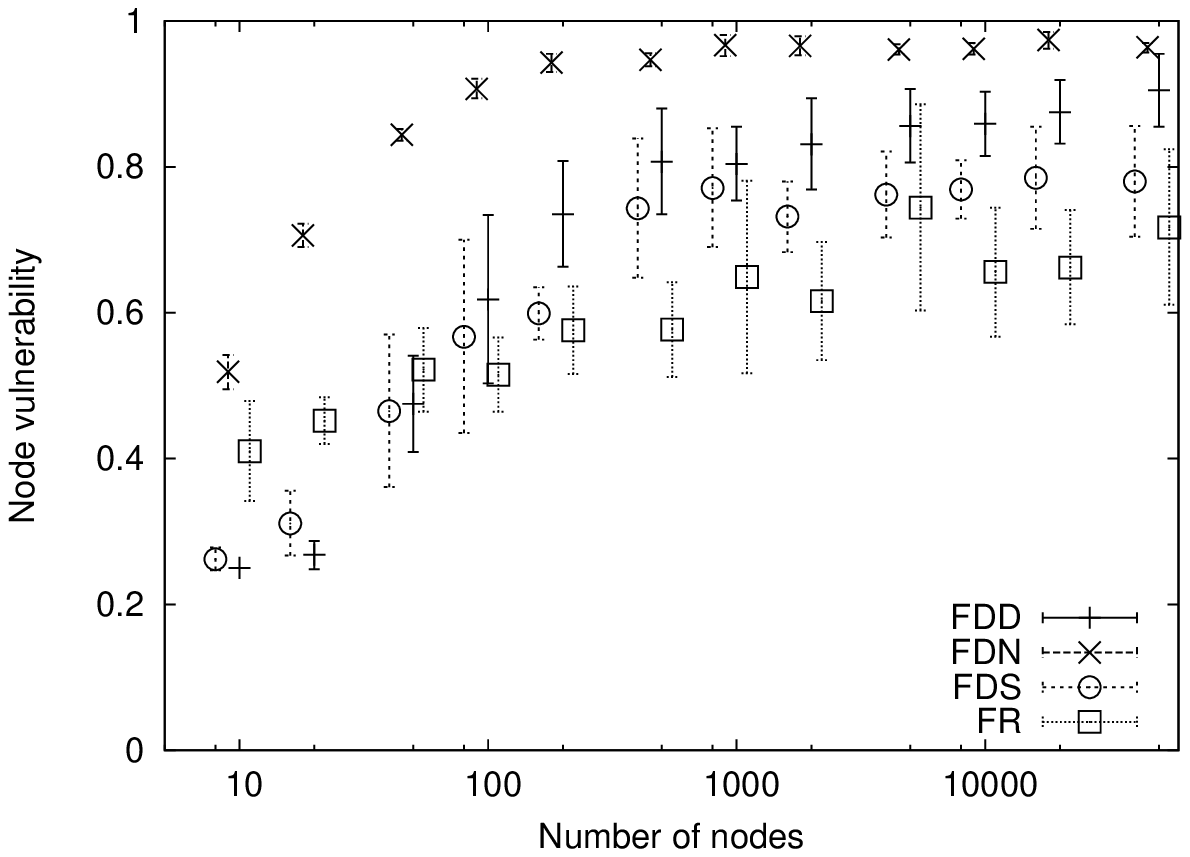}
\includegraphics[width=6cm]{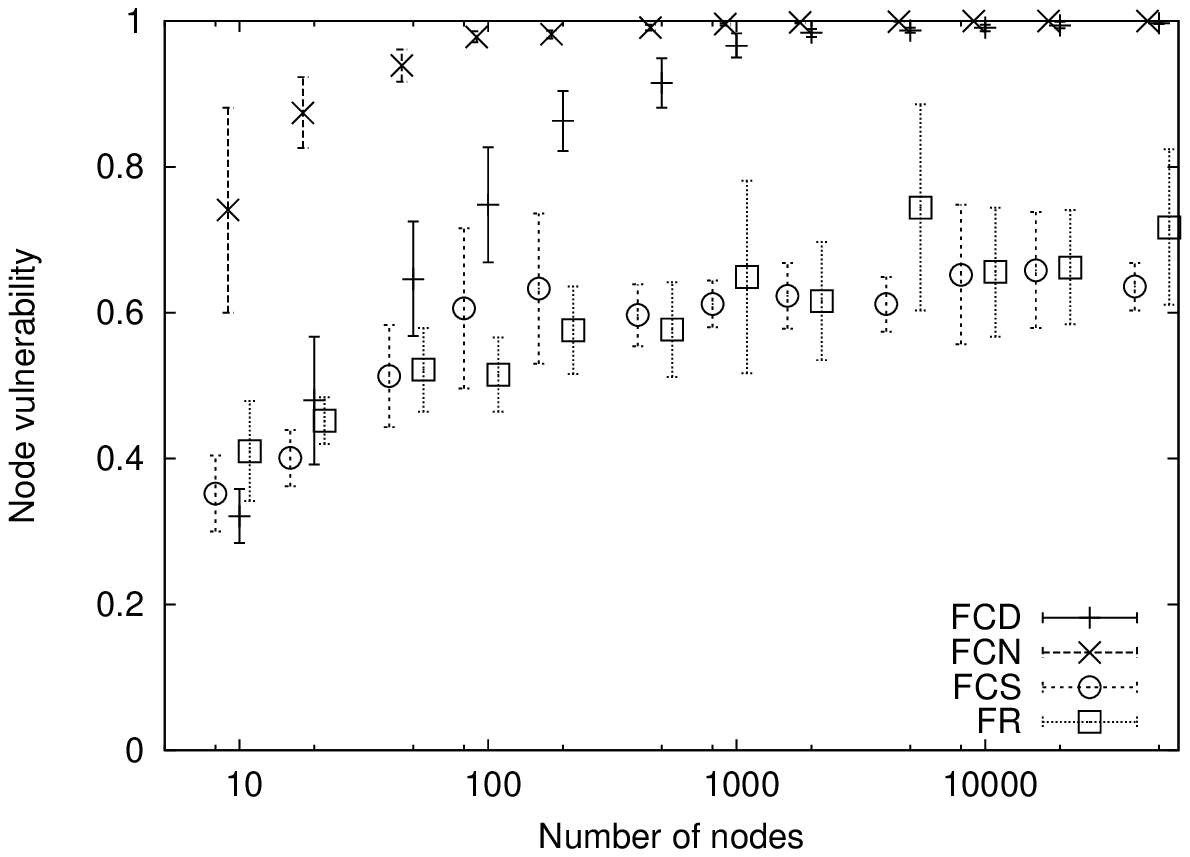}\newline
\includegraphics[width=6cm]{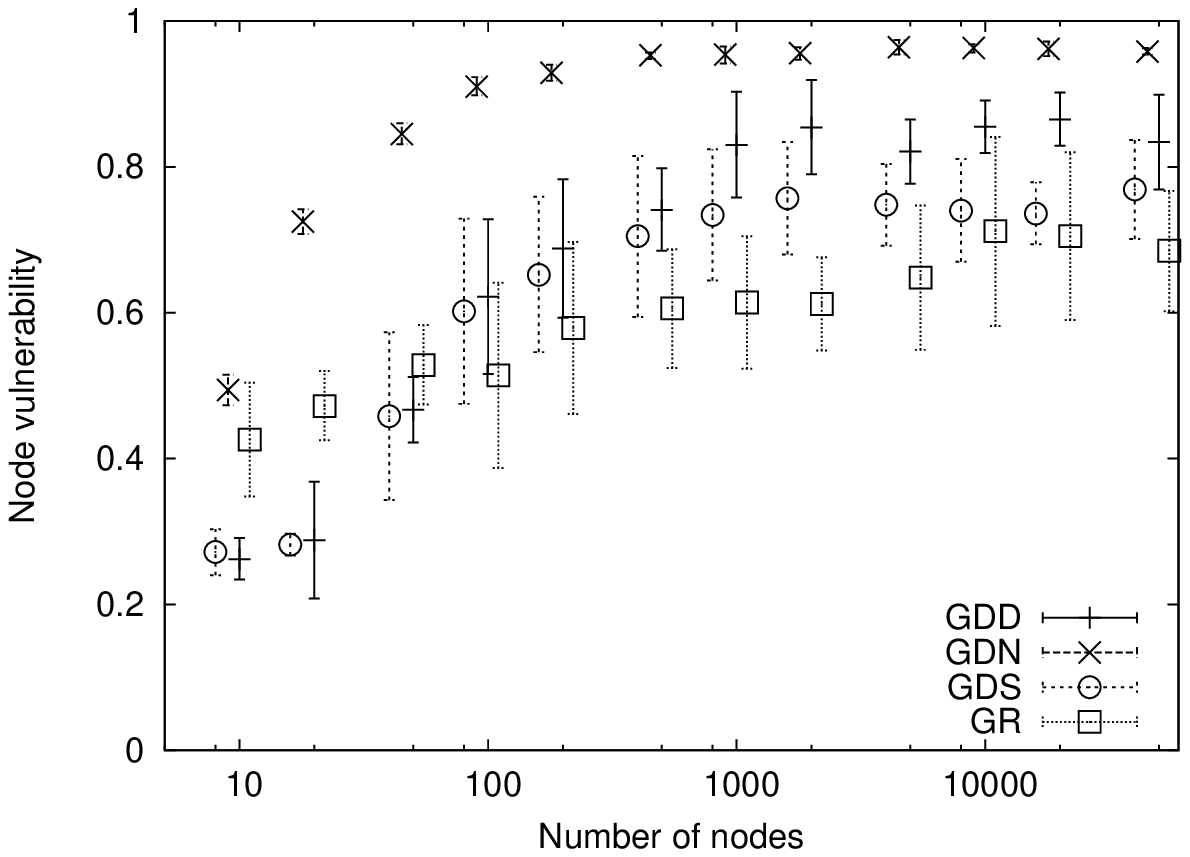}
\includegraphics[width=6cm]{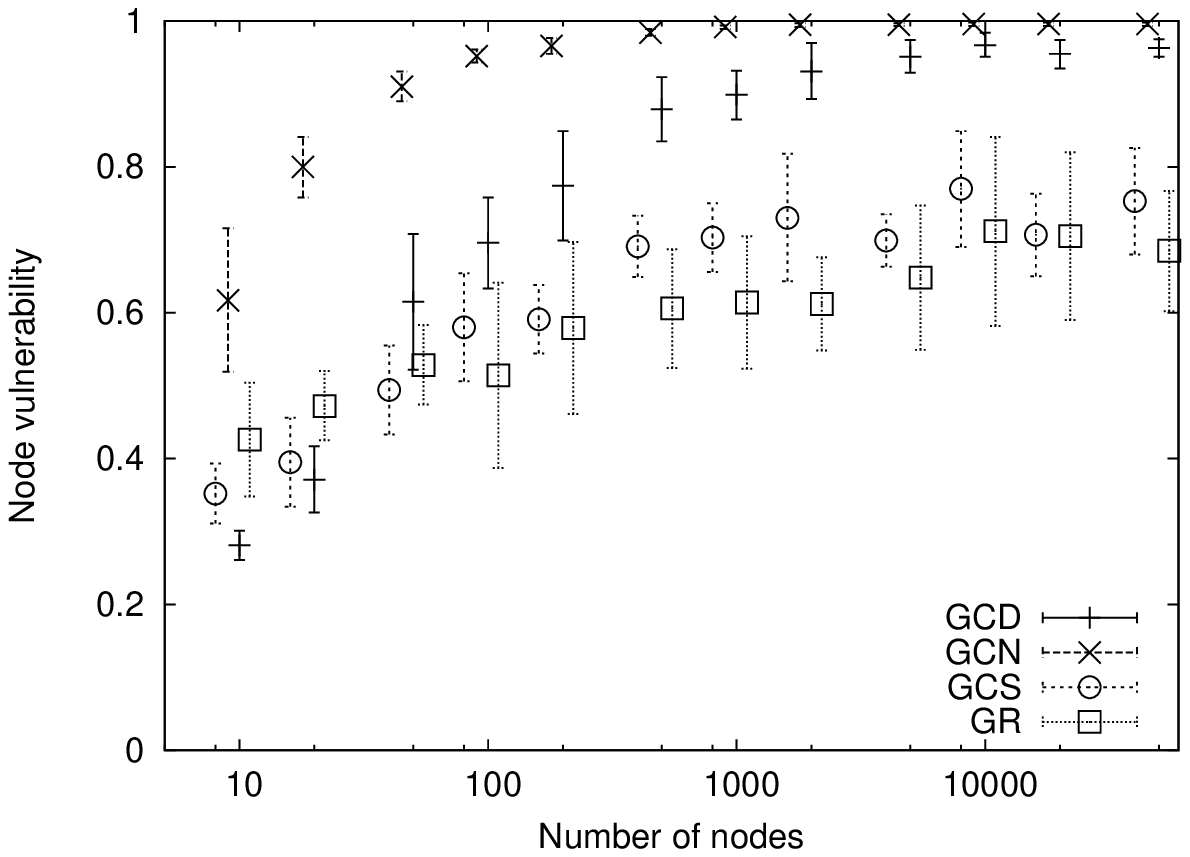}\newline
\caption{Mean node vulnerability for the various topologies and all node distributions.}
\label{fig:nvuln}
\end{center}
\end{figure}

\begin{figure}
\begin{center}
\includegraphics[width=6cm]{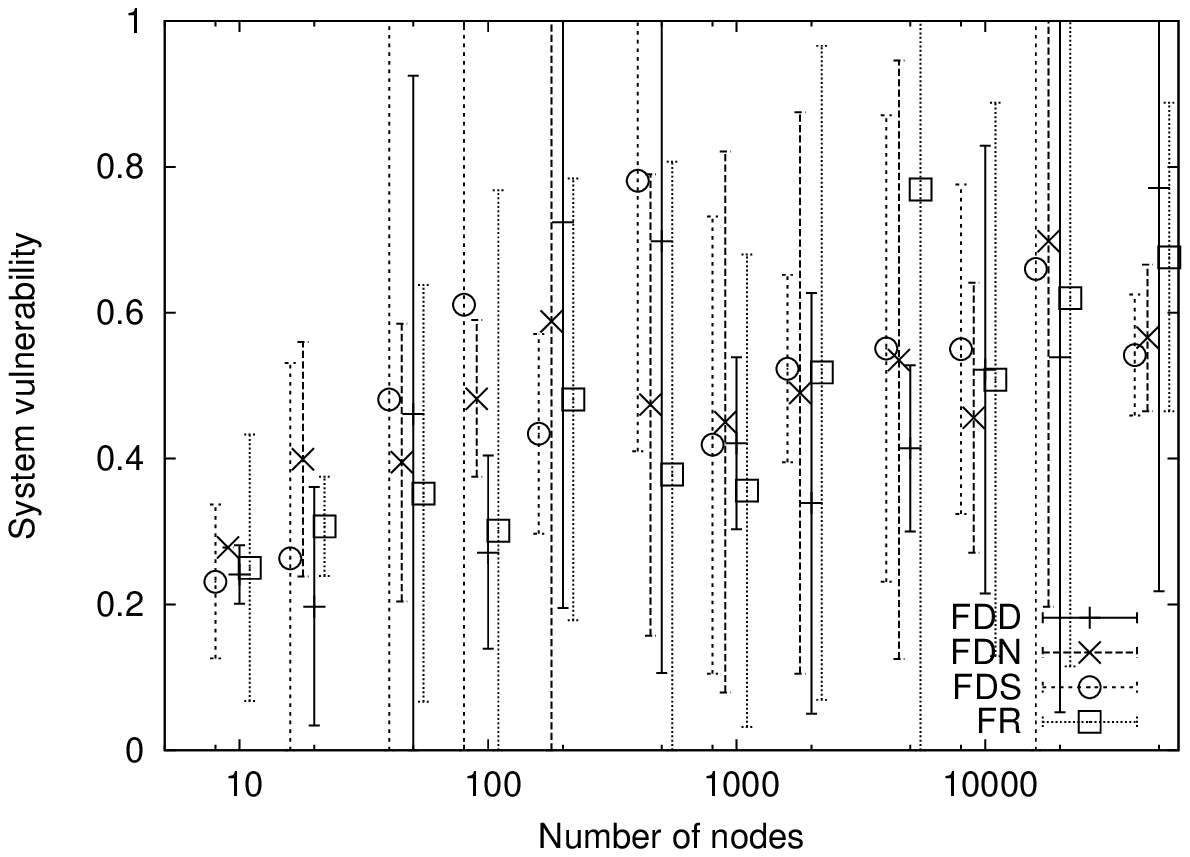}
\includegraphics[width=6cm]{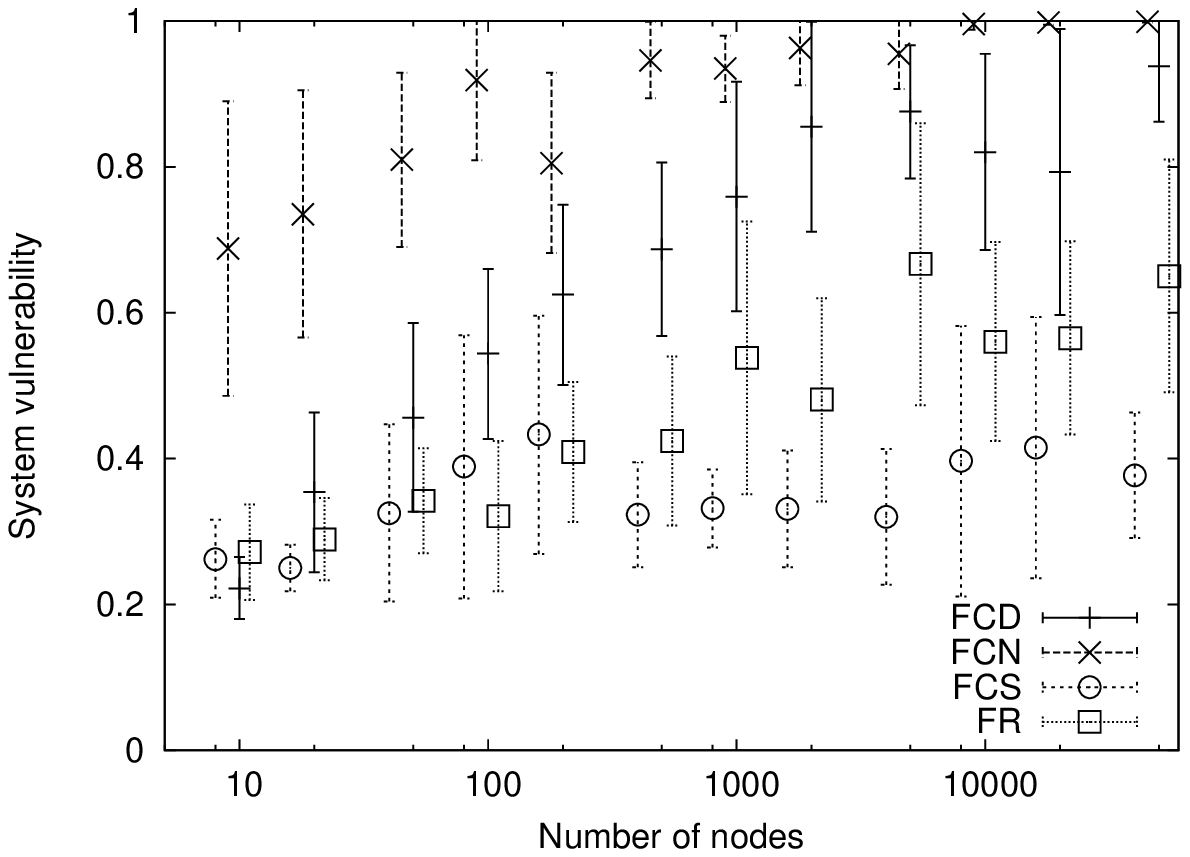}\newline
\includegraphics[width=6cm]{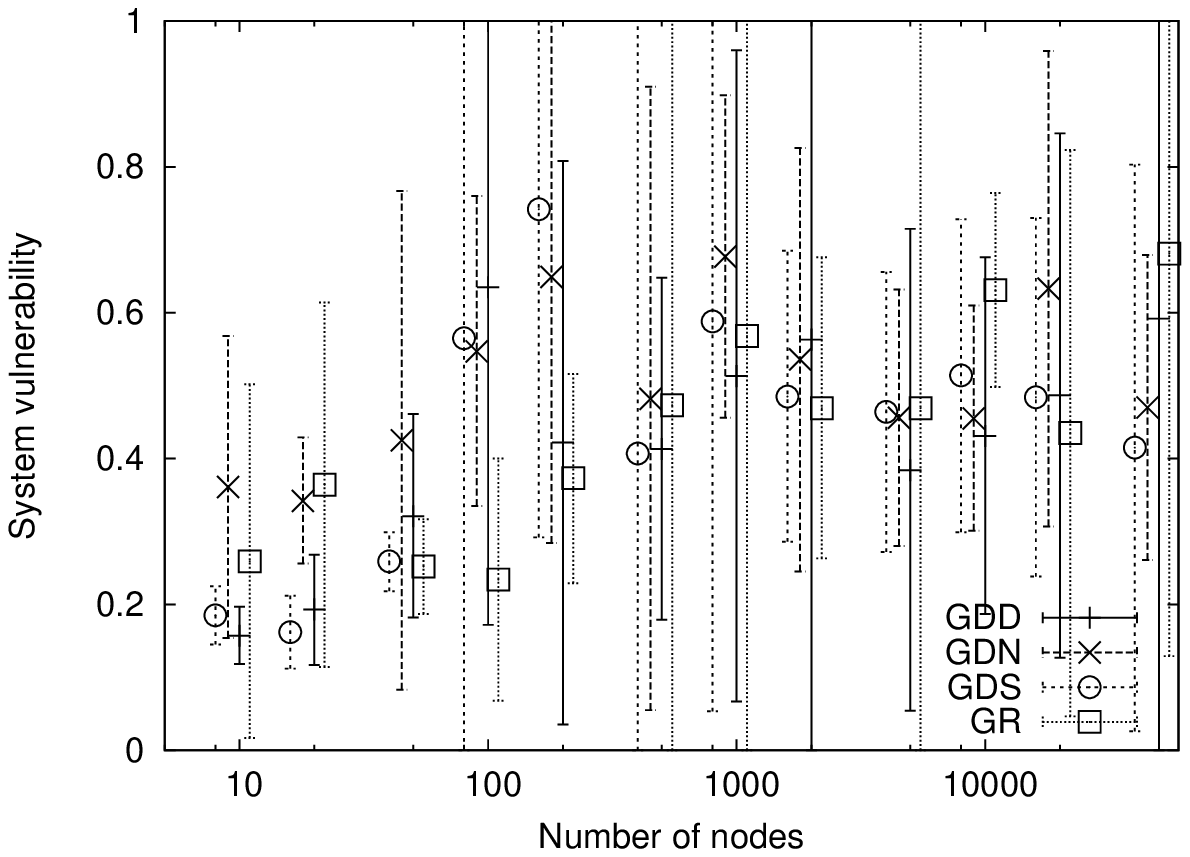}
\includegraphics[width=6cm]{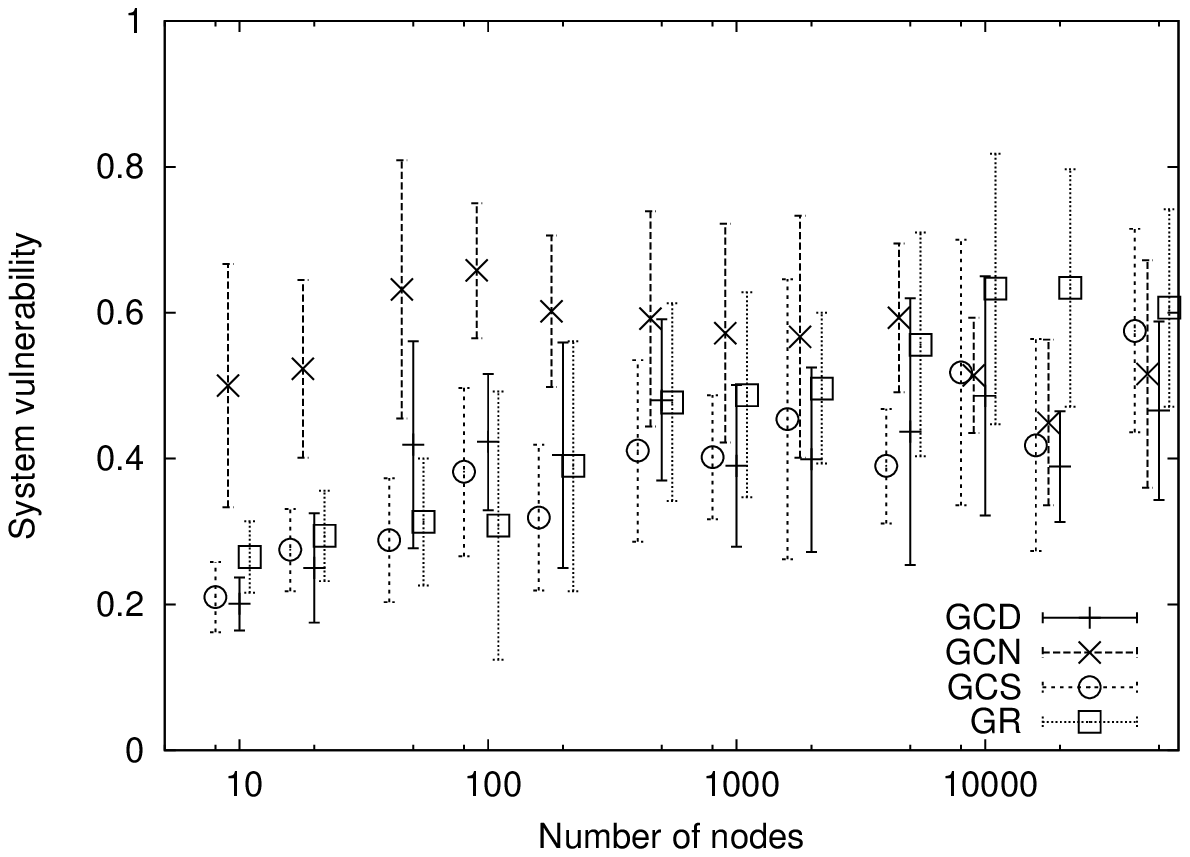}\newline
\caption{Maximum system vulnerability for the various topologies and all node distributions.}
\label{fig:svuln}
\end{center}
\end{figure}

\begin{figure}
\begin{center}
\includegraphics[width=6cm]{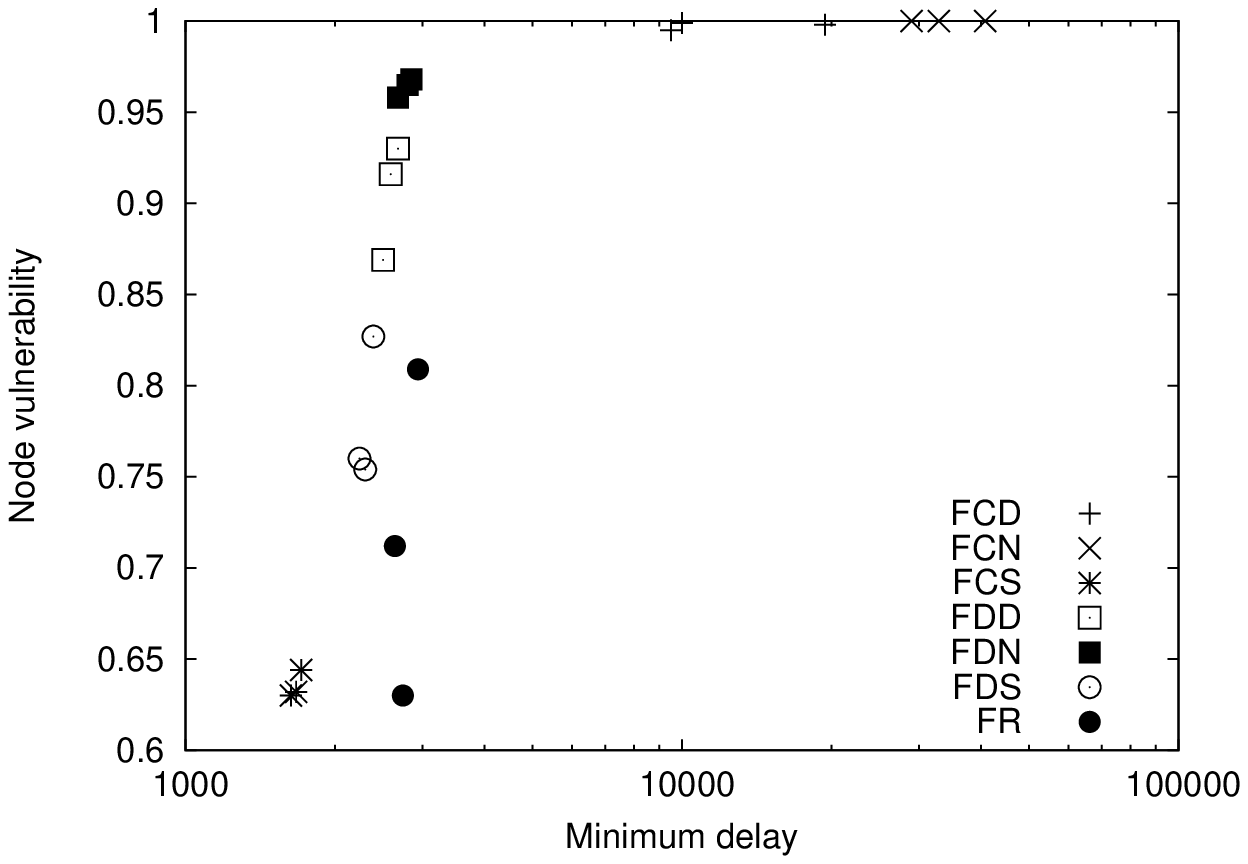}
\includegraphics[width=6cm]{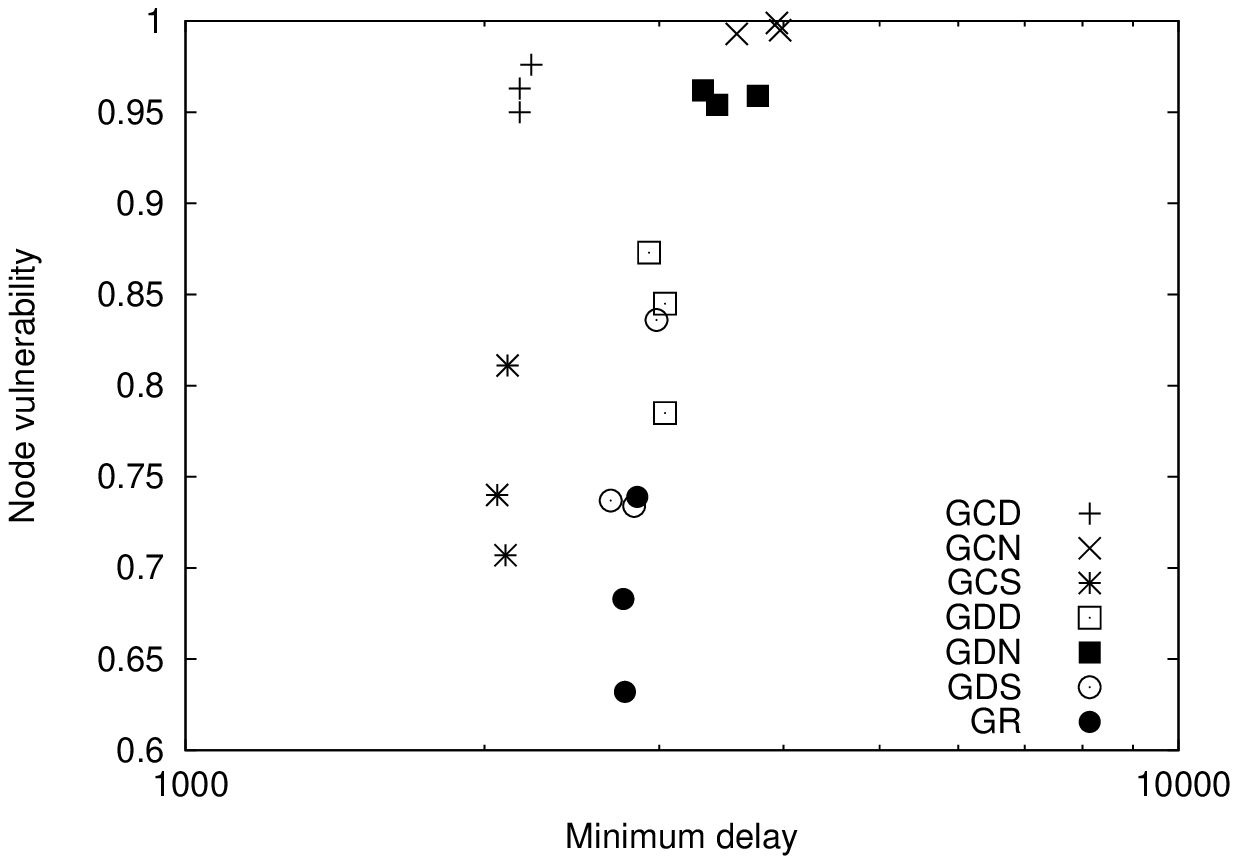}
\caption{Node vulnerability versus minimum delay for all node distributions.}
\label{fig:del_nvuln}
\end{center}
\end{figure}

\begin{figure}
\begin{center}
\includegraphics[width=6cm]{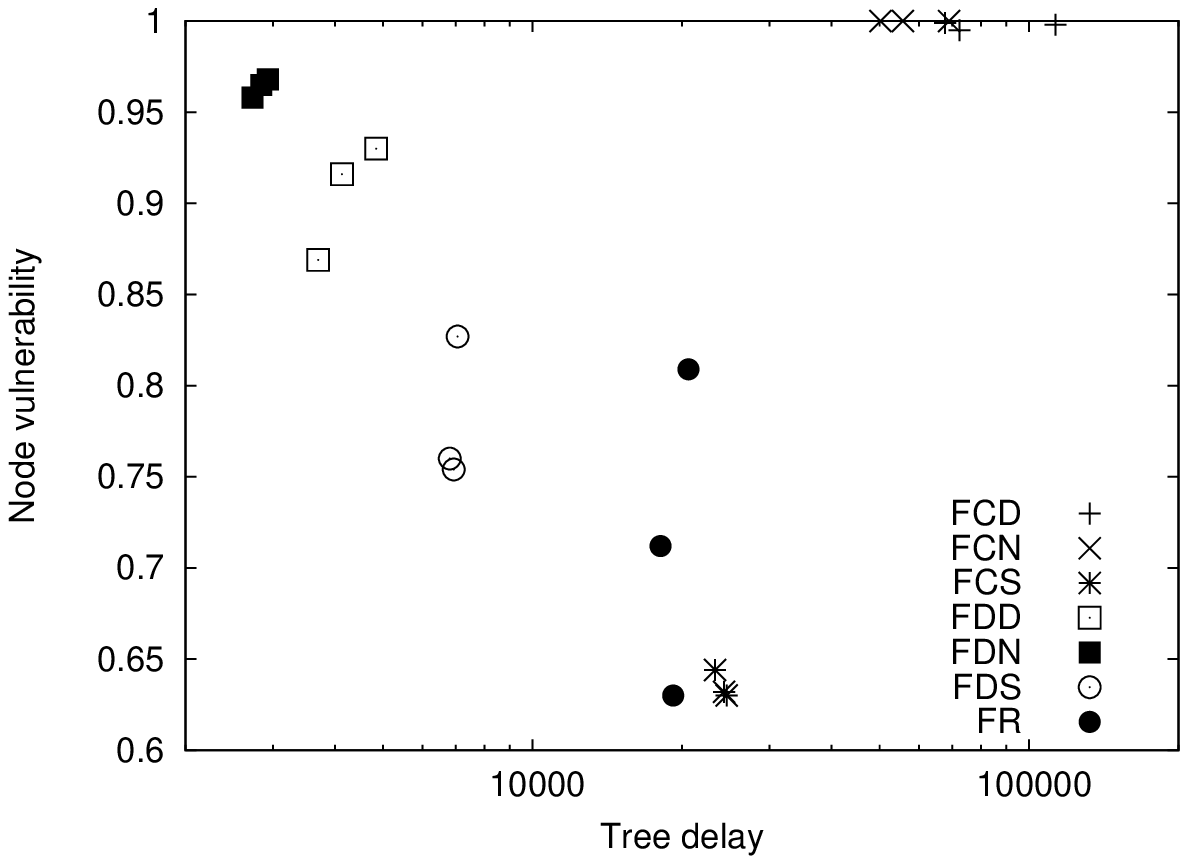}
\includegraphics[width=6cm]{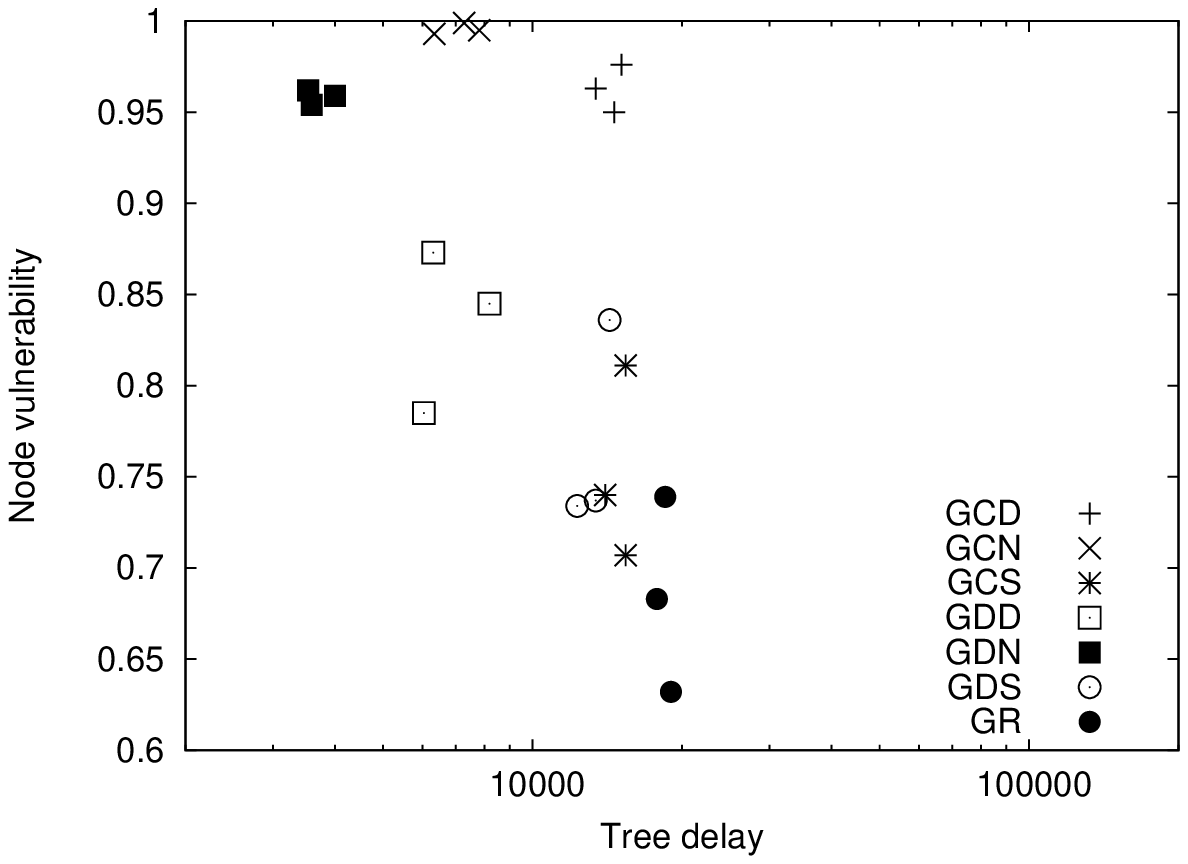}
\caption{Node vulnerability versus tree delay for all node distributions.}
\label{fig:tree_nvuln}
\end{center}
\end{figure}

\end{document}

%% file: introduction.tex
This paper investigates the impact of the topology of overlay networks on 
performance metrics for peer-to-peer live
streaming. An overlay network is a conceptual network of peers which exists on top of
the standard Internet. Peers on the overlay network
connect
according to given rules to form a topology.  There
has been recent research interest in making overlay networks
locality-aware so that peers may 
more easily find ``nearby" peers. In this 
paper we undertake a systematic evaluation of a number of
alternative locality-aware topology construction methods (and some random methods
for comparison).

The situation considered is that of a single node, known as the {\em
peercaster\/}
wishing to distribute live streaming content through a peer-to-peer network.
The peers in the network wish to download this content reliably and with a
low delay between the peercaster and themselves.
The challenge of distributing live content is somewhat different
to that of distributing recorded content on demand.  A major difference is that
delay is important to optimise (so that peers can view
streams as ``live" as possible) whereas throughput only needs
to be large enough to view the stream (a peer cannot continue
to download at faster than the rate the stream is broadcast). 

A number of strategies might be considered for forming such topologies for live streaming.
Minimising delay to the peercaster might be one strategy.  Connecting
to close (in terms of delay) nodes might be a related strategy.  Another
aspect to consider is whether it is important to aggressively minimise
delay or closeness by making as many connections as possible to
the lowest delay/closest node or whether it might be preferable
to have a range of connections. The topologies formed are tested
against several metrics which attempt to assess whether the topology is
good at reducing delay, resilient in the case of peers dropping out and
whether it ensures that the bandwidth is used fairly.

\subsection{Background and related work}

Distributing content over an overlay network has been the subject of
numerous studies in recent years.  Most of this research has been
concentrated on non-live content where the emphasis is on increasing
throughput rather than reducing delay.
In early approaches like SpreadIt \cite{spreadit}, a multicast tree is
built by centralised logic running at the data source.
Upon the arrival of a new node, the source is contacted to appoint an
unsaturated node to be the parent of the new node.
When the {\em smart-placement \/} policy is in effect, the parent node is
also selected to be close to the new node, where proximity is inferred with
traceroute messages.
More recently, Bos \cite{4410965} proposed a method which constructs a data
distribution tree containing the {\em Euclidean Minimum Spanning Tree\/},
where the distance in the Euclidean space represents the network delay.
A subset of stable and high capacity nodes are elected to become {\em super
peers\/}.
Super peers are interconnected to form a {\em Yao graph\/}, a structure
which contains the Euclidean Minimum Spanning Tree.
Normal peers attach directly to the closest super peer.
The source routed multicast tree is built over the super peers topology
based on the compass routing protocol.

The departure of a node in  a single distribution tree results in
complete loss of connectivity for all the nodes in the underlying subtree.
To overcome this problem, several studies investigate streaming the data
over a forest of multicast trees, each of which carries only part of the
stream.
CoopNet \cite{padmanabhan02distributing} is a forest-based streaming
approach, where the authors identify a tradeoff between efficiency in terms
of locality and path diversity required for resilience to node departures.
Upon addition of a new node, the source returns a significantly large set
of candidate parent nodes to ensure diversity.
As an optimisation the candidate parent nodes are selected, similarly to
SpreadIt, so that are they are {\em nearby\/} the newly added node.

Techniques for constructing trees typically assume global knowledge and at
least one interaction with the source.
Alternatively, overlay topologies can be constructed with local knowledge,
where the connections are determined by each node and the data flow may
take many alternative and potentially overlapping paths.
In \cite{ratnasamy02topologicallyaware} a technique for clustering nodes to
bins based on their locality is proposed.
As a case study of this technique, the {\em BinShort-Long\/} overlay
construction method is presented, where each node connects to $k/2$
randomly selected nodes from within its cluster (bin) and $k/2$ random
nodes from anywhere in the system.
A similar technique is proposed in \cite{1648853} as an improvement for the
BitTorrent protocol.
The clustering here is done primarily to distinguish between nodes located
in the same ISP, and nodes in different ISPs.
Out of the total BitTorrent peers discovered by a new peer through the
local tracker, all but $k$ are selected to be local peers, with typical
values 35 for total peers and 1 for the $k$ external peers.
This is done to reduce the traffic over the inter-domain links while still
maintaining enough connections with external peers to receive the data.
Finally, in \cite{4509755} the authors formulate the {\em Minimum Delay
Mesh problem \/} and prove that it is NP-hard.
They propose a heuristic for constructing a shallow (low number of hops)
and locality-aware (low delay at each hop) overlay topology.
In order to minimise the number of hops, nodes with higher capacity need to
be connected closer to the source.
The selection of the nodes to establish connections with, is done after
calculating the {\em power \/} of each node, as a function of the node's
locality and bandwidth availability.

%% file: method.tex
In order to make the simulation of the overlay tractable it is necessary to
abstract away the network itself and simulate only the overlay.  The simulation
described here makes as few assumptions as possible. 
It is assumed that each node has a fixed delay to every other node in the overlay
(as described in the next section).  It is also assumed that each node has a
sufficient download bandwidth to obtain the entire stream and upload bandwidth
to deliver a fixed proportion (which may be more than unity) of the stream.

\subsection{Node distributions}
\label{sec:locality}

Synthetic coordinate systems associate a coordinate with each peer 
in an overlay network, in such a way that the distance between the 
coordinates is a good estimate of some network property measured 
between the peers, predominantly {\em round trip time\/} (RTT). 
This can be achieved efficiently by using a limited set of 
end-to-end measurements to extrapolate those distances between 
nodes that were not explicitly measured. Thus, synthetic coordinate
systems use a limited set of measurements to model the structural 
properties of the Internet, and then use this model to predict 
end-to-end properties (such as RTT) between arbitrary peers.

The first step in the operation of a network coordinates system 
is generating a {\em distance graph\/}, where links between peers 
represent distance measurements. This distance graph is then 
{\em embedded\/} onto a space that integrates some of the 
structural properties of the Internet. Examples of these include 
a standard Euclidean space  \cite{vivaldi}, a Euclidean space 
augmented with a purely additive coordinate \cite{vivaldiSigcomm} 
or a hyperbolic space \cite{shavitt04curvature}. The 
embedding process can be viewed as an error minimization procedure 
where nodes are positioned in the space in such a way that the 
cumulative difference between the measurements and the embedded 
distances is minimized. Once this embedding has been done, and to 
the extent that the embedding space faithfully recovers the structure 
of the Internet for the measure in question, geodesic distances over 
this space are good predictors of the actual distances over the 
Internet \cite{networkCoordinatesWild}.  This space will be referred
to as {\em delay space\/}.

In the case of the simple simulation used in this paper, a standard
two-dimension Euclidean delay space is used.
Let $N$ be the number of nodes in the system excluding the peercaster.  
The $N+1$ nodes, numbered from $0$ (the peercaster)
to $N$ are distributed over the two-dimensional 
Euclidean space.  
Each node has a co-ordinate $(x_i,y_i)$ and the delay from
node $i$ to node $j$ is obtained using the standard Euclidean distance
from $(x_i,y_i)$ to $(x_j, y_j)$.

The next question for the simulation
is how to distribute the nodes on the delay space.
For the purposes of this paper we use three generation methods to create
random node distributions.  In reality, nodes in an overlay network will
cluster to some degree, for example, nodes in the real Internet are more
prevalent in some areas of the world than others (clusters in large
cities, particularly large cities with high levels of Internet usage).
In the case of an overlay network based upon nodes wishing to download
particular streaming content, the distribution will be further complicated
by whether the content is of regional, national or global interest as well
as what language the broadcast is in.  For this reason the simulation here
is tested against different assumptions about how nodes might be randomly
situated in delay space.

{\bf Flat node distribution}: In this distribution the nodes 
are flatly distributed
in a square delay space.
For each node $i$, $x_i$ and $y_i$ are chosen randomly from a flat distribution
in the interval $(-D,D)$.  In the simulations given here $D=0.25$ seconds 
(so the maximum delay between any two nodes is $\sqrt{2}/2$ secs).

{\bf Tightly clustered node distribution}: This distribution
simulates a situation where nodes are grouped into tight clusters.  The
following procedure is followed until sufficient nodes have been generated.
\begin{enumerate}
\item Coordinates position $X,Y$ is chosen with a flat distribution where
$X$ and $Y$ are chosen from the interval $(-D,D)$.  
\item The position $(X,Y)$ 
is modified by a small random perturbation $(d_X, d_Y)$ where
$d_X$ and $d_Y$ are chosen with a flat distribution in the interval 
$(-d,d)$.  
\item Coordinate $(X,Y)$ is recorded.
\item With probability $p$ go to step 1, otherwise go to step 2.
\end{enumerate}
In this distribution $D=0.25$, $d=0.005$ and $p= 0.01$.

{\bf Loosely clustered node distribution}: This distribution
is identical to the previous one but the clusters are more diffuse but on 
average contain the same number of nodes:
$D=0.25$, $d= 0.05$ and $p= 0.01$.

In each of the last two cases, after the distribution is created, the node
order is randomised.  Node order is important for local topology schemes
(see section \ref{sec:toppolicy}).  This reordering prevents nodes being
created in a convenient ``by cluster" order with nodes locally close being
created together.

\subsection{Modelling assumptions}

For simplicity it is assumed that each node attempts to download a
stream as $M$ separate and equally sized {\em substreams\/} -- note, however, 
that this could also be thought of as simply an abstraction of,
say, a chunk-based swarming system with $M$ partners from whom
equal amounts are downloaded.  Assume that
each node has capacity to download all $M$ substreams and that nodes
have upload capacities to upload only a limited number of substreams.

Each node has associated with it an upload capacity $u_i \in \mz^+$ 
which is the number
of substreams it can support (for the purposes of bandwidth calculation each
substream is considered to have a bandwidth of 1Mb/s -- although the precise unit
is unimportant and of the metrics described, only the bandwidth variance is
affected by this).  
Note that it must be assumed that
$u_0 \geq M$ (in order that all $M$ substreams can be uploaded from the
peercaster itself) and also for the system to scale it is important
that $\overline{u_i} \geq M$ (the average peer has sufficient capacity to
upload all $M$ substreams).  This is discussed more fully in section
\ref{sec:toppolicy}.
An implicit assumption is
that system bottlenecks are only at the peers in the network -- if a peer with sufficient
upload transmits to a peer with sufficient download then no intermediate link in the
internet itself will reduce this capacity.  This may not
always be the case in reality (for example several peers who belong to the same ISP may
share access network capacity in the underlying network).

Nodes will then attempt to connect to at most $M$ other peers in order
to download the complete stream (nodes can download all $M$ substreams from a
single partner node).  A node $i$ will accept at most $u_i$ connections
and request up to $M$ connections.  The complete set of connections will
be referred to as a {\em topology\/} on the overlay network.  This will
be described in the next section.

For this paper $u_i$ will be chosen from a random distribution.  In
addition $u_0$ will be fixed since it has such an important role in the 
network (naturally it must be the case that $u_0 \geq M$.  
The values used are $M=4$ and $u_i$ is chosen
with equal probability from the set $\{1,5,10,16\}$ -- in this simulation no
nodes are complete free-riders although some nodes can only produce 1/4 of
a complete stream.  The mean value of $u_i$ is $8$ so the system easily
has capacity for every node to download the stream.  As previously stated $u_0$ 
is a critical parameter in the system so $u_0 = 16$ for all simulations -- the
peercaster is always assumed to have a reasonable amount of bandwidth.  This is to prevent the simulation results being greatly dependent on this single random selection 
(a simulation where $u_0 = 1$ might get very different results from one 
with $u_0 = 16$ even if all else was the same).

%% file: topologies.tex
For the purposes of this paper a topology is defined as a graph of the connections
in the peer-to-peer network annotated with the number of connections between each
pair of peers and the upload bandwidth of each peer.  No peer is ``special" apart
from the peercaster.  The peers have no characteristics apart from an upload
bandwidth and a position which gives rise to a fixed delay between each pair of
peers.

\subsection{Topology definitions}

\begin{definition} \label{defn:feasible}
A {\em feasible topology\/} is one where
\begin{enumerate}
\item all peers have $M$ connections from which they download,
\item no peers exceed their upload bandwidth,
\item all peers can find $M$ edge distinct paths from the peercaster to themselves.
\end{enumerate}
A {\em feasible connection policy\/} is a policy for making connections which,
if followed repeatedly, will connect a set of nodes into a topology which obeys
the conditions above.  A {\em feasible connection\/} is a connection made
according to a feasible connection policy.
\end{definition}

\begin{remark}
Requirement 3 arises because it is necessary to ensure that, for example, in a
substreaming system each peer can download $M$ substreams from the peercaster.
This requirement is equivalent (by the max-flow/min-cut theorem) to requiring
that the minimum cut set to cut each peer from the peercaster is at least $M$ 
edges.
Without this requirement, a policy where node A and node B each send $M$ substreams
to the other and neither connect to the peercaster would be a feasible topology. 
\end{remark}

\begin{definition}\label{defn:thisfeasible}
The feasible connection policy used in this paper is as follows.
\begin{enumerate}
\item Initialise the system assuming only the peercaster is connected.  Let
$F:= u_0 - M$ be the spare upload bandwidth which will remain in the system
after the next peer joins.
\item Choose a peer $i$ which has $u_i$ 
such that $u_i + F \geq M$.  The choice is made according to some topology
policy (see next section).   This  guarantees that the system will have sufficient 
free bandwidth to make all $M$ connections required by the next peer.
\item Make all $M$ upload connections to peer $i$ from already connected peers (with
remaining upload capacity) according to some topology
policy (see next section).
\item Let $F:=F + u_i -M$.
\item If more peers remain to be connected then go to step (2) above.
\end{enumerate}
\end{definition}

It is easy to show that this policy will meet the requirements of
definition \ref{defn:feasible}.  Steps (2) and (4) ensure that requirement (2)
is met by checking that the new peer has sufficient upload bandwidth.  
Step (3) ensures that requirement (1) is met.

Requirement (3) must be met by step (3) of the algorithm.  The proof is
by induction.  Requirement (3) is clearly satisfied when only the peercaster
is connected.
Assume that requirement (3)
is true of the first $n$ peers to be connected.  When the next peer $n+1$
is connected by step (3) then each of the peers connected has $M$ distinct edge
paths to it.  Is it possible to form $M$ edge distinct paths
to node $n+1$?  If this were not the case then there must be some cut-set with less
than $M$ members between the peercaster and node $n+1$.  Let $U_i$ ($i \in \{1,\ldots,M\}$)
be the set of uploaders to $n+1$.  It is impossible to cut the connection to any of
the $U_i$ by removing fewer than $M$ edges by the induction hypothesis.  By construction
there must be 
exactly $M$
connections between nodes in the set $\{U_1, \ldots, U_M\}$ 
and node $n+1$ so to cut between this set
and $n+1$ obviously all $M$ connections
would need to be removed.  No cut set of less than $M$ members exists between the
peercaster and $n+1$ exists and hence requirement (3) of definition \ref{defn:feasible}
is met.

\subsection{Topology policies}
\label{sec:toppolicy}

In this paper a {\em fixed\/} policy is one where the whole ``universe" of
peers is available from the start and connections can choose from this
universe.  Conversely, a {\em growing\/} policy is one where peers 
arrive one by one and each peer makes all its connects when it arrives.
In earlier work on this subject 
\cite{ukpew} 
the terms global and local were used instead.  
A topology which connects {\em closest\/} peers is one which chooses 
the {\em feasible connection\/} which has least delay between the two
peers being connected.  A topology which
connects {\em least delay\/} peers is one which chooses the
{\em feasible connection\/} which has the smallest value for the
shortest delay path from the peercaster to the peer on the download
end of the new connection.

\begin{remark}
The real difference between fixed and growing topologies is that a growing 
topology connects nodes in the order in which they appear.  A fixed topology
is allowed to choose which node to connect.  In theory, a fixed
topology has much more freedom and could perform much better.
\end{remark}

In this paper {\em connection diversity\/} refers to topologies which attempt to upload 
from distinct peers wherever possible.  If a topology naively selects the {\em closest\/}
peer for example then it is likely to make multiple connections to the same peer (indeed 
this will happen unless that peer has its upload bandwidth exhausted).  With {\em
connection diversity\/} then a peer will have more than one connection to the same uploader
if and only if no other connection is available.  A {\em small world\/} topology is one which 
makes $N-1$ connections with connection diversity and the final connection completely at random.

The policies for the fixed topologies are as follows.
\begin{itemize}
\item FR -- Fixed random.
\item FCD -- Fixed closest, with connection diversity.
\item FCN -- Fixed closest, no diversity.
\item FCS -- Fixed closest, small world.
\item FDD -- Fixed least delay, with connection diversity.
\item FDN -- Fixed least delay, no diversity.
\item FDS -- Fixed least delay, small world.
\end{itemize}

GR, GCD, GCN, GCS, GDD, GDN and GDS are the equivalent topologies for the
``growing" peer sets.

It will help the reader's understanding to describe two of these policies more fully.
The policy GR (growing random) is implemented using definition \ref{defn:thisfeasible}
as follows.  In step (2) of the policy, only one peer (call it peer $i$) 
is available at a time
and therefore this choice is fixed.  In step (3) of the policy, a random peer is
chosen from the set of peers which are already connected and which have spare
upload capacity.  This peer is connected as an uploader to peer $i$ and its upload
capacity is reduced accordingly.  This is repeated $M$ times.

The policy FDD is implemented using definition \ref{defn:thisfeasible} as follows.
Let $d_j$ be the shortest path delay from the peercaster
to node $j$ or $\infty$ if node $j$ is not
yet connected.  Let $d(i,j)$ be the delay from peer $i$ to peer $j$.
In step (2) of the policy, the peer $i$ chosen is the peer with the
smallest value for $d_j + d(i,j)$ which has a sufficiently large $u_i$ to meet the
condition of step (2) (note that $u_i = 0$ is large enough if $F = M$).  It is now
necessary to make $M$ connections (with diversity) to peer $i$.  This is achieved
by connecting to the peer with the smallest value of $d_j + d(i,j)$ and then setting
$d_j:= d_j + L$ where $L$ is some ``large" number\footnote{$L$ should be large
enough that a second connection to $j$ will only be made if no non-penalised node is
available -- $N \max(d(i,j))$ is sufficient.}.  This is repeated until $M$ connections
are made.

\begin{remark}
It should be noticed that in the FR topology the nodes are selected in 
a random order and it is, therefore, effectively the same as the GR topology.
\end{remark}

\subsection{Metrics for topologies}

Because each node has $M$ independent connections, variants on more 
usual network metrics are used here.  For example, it is not 
simply the shortest path
from a node to the peercaster to the node which is of interest but the 
path length along all paths.

The metrics listed in this session have been created with several
considerations in mind.  A ``good" topology should have all or
most of the following properties.
\begin{itemize}
\item Low delay to end nodes -- this translates to nodes being able to
view streams with good ``liveness".
\item High resilience to churn -- a peer-to-peer network is, by its
nature, highly dynamic.  The loss of any single node should
not greatly affect the network.  
\item Diversity of paths -- related to the above,
an individual peer would want a diverse
set of connections so that the loss of a single intermediate 
node will not affect every substream it is downloading.
\end{itemize}

Let $D_k(i)$ be the shortest path from the node $i$ to the peercaster if the
first hop is the $k$th uploader to node $i$.

\begin{definition}
The {\em minimum delay\/} of a node is the shortest path distance from the 
peercaster to the node -- it is the minimum over $k$ of $D_k(i)$.  
The minimum delay of a system is the mean of this 
taken over all nodes.  This metric gives an estimate of the minimum possible
end-to-end liveness that any of the substreams a node gets will experience.
\end{definition}

\begin{definition}
The {\em tree delay\/} of a node is the distance from the peercaster to
the node after the removal of $M-1$ peercasted rooted shortest
path trees (that is to say,
remove the shortest path from the peercaster to each peer and repeat
this operation $M-1$ times).  This metric
estimates a pessimistic end-to-end delay if packets took an extremely
favourable path.
The tree delay of the system is the mean of this taken over all nodes.
\end{definition}

Let $V(i)$ be
the substreams connecting node $i$ to the peercaster which could potentially 
be disrupted by the removal of a single node (not including the
peercaster).  It is zero if and only if every node 
is directly connected to the peercaster.  It is $M$ if all of the paths
$D_k(i)$ go through a single node (that is every path to $i$ could be
cut by the removal of a single node).

\begin{definition}
The {\em mean node vulnerability\/} for the
system is $\sum_{i=1}^N (V_i)/(NM)$ -- this
is one if every node has a single node which could cut all $M$ substreams 
(every node is vulnerable to the loss of all substreams) and
is zero if every node has all connections directly to the peercaster
(every node cannot be cut off).
\end{definition}

Let $S_i$ be the vulnerability of the system to the removal of node $i$.
It is, in a sense, the dual of $V_i$.  It is the total number of streams
$D_j(k)$ (where $j \neq i$) which could be broken if node $i$ were
removed from the system.

\begin{definition}
The {\em maximum system vulnerability\/} is
given by $\max_i S_i/(NM)$ -- 
this is the proportion of paths which could potentially be damaged by the 
removal of a
single node.  It will be one if there is a single node (apart from the peercaster)
which can disrupt every transmission path and zero if there are no nodes which
can damage paths (only possible if every node connects directly to the peercaster).
This measure is similar to finding the node with maximum
Betweenness-Centrality \cite{betweenness}.  It is a measure of the worst case
damage a single peer leaving the network can cause (one meaning a single peer leaving
can disconnect every path and zero meaning no peer leaving can disconnect any paths).
\end{definition}

%% file: results.tex
Each topology is run three times for each of the three node distributions 
and for each of the fourteen topology algorithms.  The algorithms are run on
10, 20, 50, 100, 200, 500, 1000, 2000, 5000, 10000, 20000 and 50000 nodes.  This
gives a total of 1,512 total simulation runs.

Figure \ref{fig:distributions} shows the effects of the node distribution algorithms.
The scale is delay in milliseconds.  Each dot on the plots represents a node in
the distribution.  The cartesian distance between any pair of points is the
delay between them.

Figures \ref{fig:del}--\ref{fig:svuln} show various metrics
versus number of nodes for each of
the topologies.  Each point on the graphs is a mean over the three runs and the
three node distributions and the error bars represent
a 95\% confidence interval.  The top left plot
shows the fixed topologies optimised by delay and the fixed random
topology.  The top right shows the fixed topologies optimised by closeness.
The bottom plots show the same for the growing topologies.  The scales 
on the graph are kept the same for each metric for ease of comparison.
The error bars are shifted slightly left and right of their true x position
to prevent them overlapping.

Figure \ref{fig:del} shows the results for minimum delay.  Somewhat surprisingly
the fixed topologies where the algorithm has a free choice of which node to
connect, have larger delays than the growing topologies where the nodes
are connected in order of arrival.  Of all the policies the small world policies
have low delay in almost all circumstances.  Those policies which do not
attempt to introduce any diversity into connections perform very poorly.  Somewhat
surprisingly many policies actually perform worse than random including many
of those which
connect using closeness not delay
and many of the policies which use no diversity.  The best policy
overall is FCS.

Figure \ref{fig:tree} shows the results for tree delay which is a measure
of the maximum likely delay for the topology.  These results are sometimes the 
opposite of those in 
Figure \ref{fig:del} in the sense that those policies with no
diversity perform well.  
In this case, the best policies are almost always those
which connect using no diversity.   The definition of tree delay
explains why these results are so different for those for minimum delay.
The tree delay definition is pessimistic about network performance
almost to the same degree that the minimum delay is optimistic.
Any long link is likely to be taken into acocunt in the calculation and
this is why
overall the random topologies perform
worst in most circumstances.  The small world topologies perform better than 
random in most cases.  In this case it is less clear whether growing topologies
are better or worse overall than random.  However, the expectation that the
extra freedom in the fixed topologies would provided better performance is
not met in general.  The best performing topology is FDN followed by GDN.

Figure \ref{fig:nvuln} shows the results for node vulnerability.  As might be
expected random and small world topologies have the lowest vulnerability
and those with no diversity have the highest vulnerability.  Figure 
\ref{fig:svuln} shows the results for system vulnerability.  The error bars on
these measurements are large showing that this measure is extremely dependent on
the precise details of the simulation.  In general it seems that the random
and small world topologies are slightly better than those with no diversity but
the high variability of the results makes it hard to say more.

Figure \ref{fig:del_nvuln} shows the minimum delay plotted against node vulnerability.
Every point on the graph represents the mean for a given topology and a given node
distribution averaged over the three runs.  As can be seen, the three points for each
topology are generally close (although vary in the y axis) indicating that the
node distribution has little effect on the minimum delay metric.  The best topology 
policies for the combination of node vulnerability and minimum delay
are FCS and GCS.  FDS and GDS also perform relatively well.

Figure \ref{fig:tree_nvuln} shows the tree delay plotted against node vulnerability.
Again the three points for each topology are close on the plots (the main exception 
being FR and GR) showing that the node distribution has little impact on the 
results in most cases.  The best policy here is less clear.  If 
system vulnerability is considered most important then FCS, FR and GR would be the
best.  If tree delay is the most important then FDN and GDN are better.
For a compromise between the two, FDS, GDS, GCS and GDD perform well.

%% file: conclusions.tex
The work presented here shows some initial results for topology creation algorithms
for peer-to-peer networking which are aware of delays between peers.  
The results here show that naive policies which connect networks according to
delays or closeness are not always successful.  Indeed, those policies often
do not perform well at all.  Overall, policies involving a random component
(the so-called small-world policies) perform well over a variety of metrics.
The results show that there is great benefit to be had in arranging topologies
according to delay.  However, they also show that naive policies to do this can
do more harm than good.

An interesting outcome of this research is that, for the parameters used here,
the system seemed extremely insensitive to the node distribution used.  The node
distribution policies were chosen so that the nodes were laid out in a delay space
of approximately the same size.  However, only for the global closest topology policy
were significant differences found in metrics due to a change in the node distribution.
This is important since, if this conclusion is more widely applicable, it could
free modellers from the (possibly extremely time consuming) task of attempting to
validate a peer-to-peer model against a realistic distribution of global delay.

There are many other simulation parameters which could be investigated.  The choice
of four substreams here and the distribution of upload capacities was somewhat
arbitrary.  However, it is difficult to run simulations with too many ``degrees
of freedom".  A repeated experiment with only one node distribution topology but
differences in the distributions of upload bandwidths might generate some interesting
results.  Indeed a large problem with this research is that the state space to explore
is extremely large even in this simple simulation. 

The metrics used here are far from perfect.  Testing the algorithms in a more detailed
peer-to-peer simulation is an obvious next step.